\documentclass[a4paper,prb,showpacs,superscriptaddress]{revtex4}    
\usepackage{epsfig}

\begin{document}

\title{Effect of Plasma Irradiation on ${\rm CdI_2}$ films.}

\author{R. S. Rawat}
 \affiliation{National Sciences, National Institute of Education, Nanyang
Technological University, Singapore}
\email{rsrawat@nie.edu.sg}
\author{P. Arun}
 \affiliation{Department of Physics \& Astrophysics, University of
Delhi, Delhi - 110 007, India}
  \email{arunp92@physics.du.ac.in, arunp92@yahoo.co.in, agni@physics.du.ac.in}
\author{A. G. Vedeshwar}
 \affiliation{Department of Physics \& Astrophysics, University of
Delhi, Delhi - 110 007, India}
\author{P. Lee}
 \affiliation{National Sciences, National Institute of Education, Nanyang
Technological University, Singapore}
\author{S. Lee}
\affiliation{International Centre for Dense Magnetised Plasma, Warsaw,
Poland.}
\begin{abstract}
The effect of plasma irradiation is studied systematically on a 4H polytype
(002) oriented ${\rm CdI_2}$ stoichiometric film having compressive residual
stress. Plasma irradiation was found to change the orientation to (110) of
the film at certain moderate irradiation distances. A linear decrease in
grain size and residual stress was observed with decreasing irradiation
distance (or increasing ion energy) consistent with both structural and
morphological observations. The direct optical energy gap ${\rm E_g}$ was
found to increase linearly at the rate ${\rm 15\mu eV/atm}$ with the
compressive stress. The combined data of present compressive stress and from
earlier reported tensile stress show a consistent trend of ${\rm E_g}$
change with stress. The iodine-iodine distance in the unit cell could be
responsible for the observed change in ${\rm E_g}$ with stress.
\end{abstract}


\maketitle

\section{Introduction}
${\rm CdI_2}$ has a layered structure with neighboring layers held by Van
der Waals forces. Different stacking sequences of iodine, sandwiching Cd
layers give rise to polytype structures. As many as 200 polytypes are
recorded, however only few occur commonly \cite{poly}. Recent studies have
revived interest in cadmium iodide films \cite{PT1, PT2, PT3, PT4}. These
reports illustrate the variation of optical properties of ${\rm CdI_2}$ film as 
a function of film parameters like its thickness, deposition rate, substrate 
temperature, effect of heat-treatment etc. These parameters were shown to
effect the grain size and residual stress in the film. Thus, the optical 
properties were shown to have a strong correlation with the grain size and
residual stress. Grain size was also shown to increase linearly with the 
residual tensile stress beyond the threshold. The growth of grain size and its
distribution, as stated, depends on parameters like film thickness,
deposition rate etc. However, residual stress in the film can also alter
film properties and performance significantly. In view of the potential 
applications in surface science\cite{Gerlach} and integrated 
electronics\cite{Skotheim}, studies on ion beam effects and plasma
processing have gained significance in recent years. Desired and controlled 
modifications of physical and surface properties have been reported on ion
irradiation\cite{Loudiana}. Material modification such as grain size,
morphology etc, can be achieved by bombarding material with radiation or 
highly kinetic ions (plasma). This
manuscript details the modification of optical properties by plasma
irradiation (highly energetic Argon ions).

\section{Experimental Details}

Films of ${\rm CdI_2}$ were grown on glass substrates at room temperature by
thermal evaporation at vacuum better than ${\rm 10^{-6}}$Torr. The starting 
material was 99.999\% pure stoichiometric powder which was pelletized before 
placing it in molybdenum boat for evaporation.  The film thickness was monitored 
during its growth by quartz crystal thickness monitor and was subsequently 
confirmed by Dektek IIA surface profiler (which uses the method of mechanical 
stylus movement on the surface). The movement of the stylus across the edge of 
the film determines the step height or the film thickness. The film
thickness was found to be uniform over an area of 6cm $\times$ 6cm. 
The structural, chemical composition, morphology and optical
absorption measurements of the films were 
carried out by X-ray diffraction (Philip PW1840 X-ray diffractometer), 
photo-electron spectroscopy (Shimadzu's ESCA 750), scanning electron 
microscope (JOEL-840) and UV-vis spectrophotometer (Shimadzu's UV-260) 
respectively.

It was found that
very slow deposition rate ($<$ 1nm/sec) led to oriented film growth with
tensile residual stress. However higher deposition rates (2-5nm/sec) lead to
films with varying cell parameter (c) with increasing film thickness. 
The nature of residual stress was determined from X-Ray diffraction data. 
Since the effect of tensile residual stress on film properties has been
extensively studied in an earlier report \cite{PT4} the films in this study 
were grown with a deposition rate $\geq$ 2nm/sec to study the effect of
compressive residual stress on film properties. Further, the plasma
processing of the films is also pursued. Small pieces of
the samples were cut and irradiated with highly kinetic Argon ions produced
by a Dense Plasma Focusing (DPF) device. 

DPF is used as a source of neutrons, 
X-rays, energetic ions, and relativistic electrons. Being such a 
versatile source, it has been used in various applications    
such as a neutron source for pulsed activation analysis \cite{1}, a spectroscopic 
source for production of highly ionized species \cite{2}, a pump source for
lasers \cite{3}, a high flux X-ray source for lithography \cite{4}, an 
electron source for micro-lithography \cite{5}, and a highly energetic ion 
source for processing of materials \cite{6, 7, 8, 9} and thin film deposition 
\cite{10, 11}. The Mather-type DPF device at NIE-SSC-PFF, Singapore was used
\cite{14}. Such a device is powered by a single 30$\mu$F, 15 kV fast 
discharging capacitor. Self-generated magnetic fields caused by the transfer of 
capacitor voltage across the electrode assembly of the DPF device result in the 
formation of highly dense and hot plasma column during the collapse phase of
plasma dynamics in the DPF device, just above the anode. 
Fig. 1 shows the position where the film sample is mounted inside the DPF
for exposure just above the anode.

An aperture assembly is kept between the film sample and the anode. This 
reduce the amount of copper debris (contributed by the copper anode) 
accumulating on the samples to be irradiated. During the entire investigation, 
a charging voltage of 14 kV was used for the DPF device. The working gas used 
was argon, which was kept at a filling gas pressure of 1 mbar. 
A shutter was placed between the aperture and the film sample to   
avoid exposing the sample to weak ion beams produced while optimizing the
DPF device for strong focusing. The shutter was removed after optimum focus 
was achieved, thereby exposing the sample to energetic ions in the next DPF
pulse. By varying the distance of the film from the 
anode, the average kinetic energy of Argon ions impinging the film can be
varied. The argon ion energies at various distances from the top of the anode 
were measured by Rawat and Srivastava \cite{16} using a biased ion collector.

\section{Experimental Results}
All the films grown at room temperature were stoichiometric and
polycrystalline. The stoichiometry was confirmed by ESCA (Electron
Spectroscopy for Chemical Analysis). It has been reported that ${\rm CdI_2}$
films grown or annealed above 300K were polycrystalline \cite{film}. The
X-ray diffractograms of as grown and plasma irradiated films are shown in
fig 2. Our X-ray diffraction data agrees quite well with the powder
diffraction data ASTM card 12-574. The structural polytype of the films can
be identified as 4H from the cell dimensions. The various diffractograms are
identified in the caption. Figure displays only few representative
diffractograms for the purpose of clarity although all the samples were
studied. The important observation is the (00l) parallel to substrate
plane oriented growth in as grown and plasma irradiated at larger distance
samples as revealed by the (00l) major peaks. Such growth of preferred
orientation of crystallite planes has been observed in earlier studies
\cite{PT1, PT2, PT3, PT4, 3.11, 3.12} as well. However, we can see a
modification in the crystallite orientation from (00l) to (hh0) by the
plasma irradiation at moderate distances. This fact can be noticed explicitly
in fig 3 where we have plotted the intensity of two major peaks changing
with plasma irradiation. From the figure, we treat three samples (irradiated
at 6, 7 and 8cm) as mainly (hh0) or (110) orientated and rest as (00l)
oriented films.

X-ray diffraction data was also used to determine the residual (or internal) 
stress in
the sample. The displacement of diffraction peaks from their corresponding 
powder data indicates a uniform stress developed in the film during 
condensation. If the diffraction peaks shift to lower angle (increasing {\sl
d}), a tensile stress can be realized. Similarly, the decrease in {\sl d} 
indicates a compressive stress \cite{61, 71}. Figure \ref{shift} shows the
displacement of the strongest peak, (002) from its corresponding powder data 
(ASTM No.12-574) as shown by the vertical broken line. 
The shift to the higher side indicates a uniform compressive stress in the
films. The identification of samples is same as in Fig (2). We adopt the same 
method used by Pankaj et al \cite{PT3, PT4} to quantify the residual stress present in the film.
The actual residual stress can be estimated by the strain produced, given by 
the expression 
\begin{eqnarray}
{\Delta d \over d} = {d(Observed) - d(ASTM) \over d(ASTM)}
\end{eqnarray}
Equation (1) determines the strain produced in d spacing of (00l) planes
which are the basal planes perpendicular to the 'c' axis of the unit cell.
The residual stress along in this direction can then simply be obtained by
multiplying the strain with the appropriate elastic constant of ${\rm
CdI_2}$. The diagonal elements (e.g. ${\rm C_{11}}$) of elastic tensor
represent pure tensile or compressive components. ${\rm C_{11}=4.91 \times
10^{10}N/m^2\, (or\, 4.85 \times 10^5 atm)}$ for ${\rm CdI_2}$
crystal\cite{data}. The stress on (110) for (110) oriented three samples was
determined similarly from the corresponding peak.

We have determined the cell parameters for all the samples studied. As
mentioned above the direct consequence of the residual stress along 'c'
axis, i.e. on (002), is the change in 'c' parameter. The 'c' parameter
determined for the plasma irradiated films at various distances from the
anode was found to vary linearly with the residual stress as shown in figure
5. The as grown film had maximum compressive stress and plasma irradiation
relaxes this stress. Good linear behavior also indicates the well oriented
film growth. However, it should be noted that we have not included three
data points of (110) orientation here although there is a little stress on
(002) as well caused by that on (110).

We have characterized all the samples for their morphology by SEM. The as
grown films were having an average grain size of ${\rm 4.8\mu m}$. We have
displayed few representative SEM pictures of plasma irradiated samples in
fig 6. Morphological manifestations seem to be consistent with the
structural studies. Morphology of samples irradiated at 7 or 8cm is quite
different from those irradiated at 9 or 10cm because of different
crystallite orientations as discussed earlier. All the samples having (002)
orientation show similar morphology but with different grain size. However,
the sample irradiated at 7cm differs slightly in morphology from that
irradiated either at 6cm or 8cm although all three samples have same
orientation (110). This could be possible because the sample irradiated at
7cm has minimum number of (002) planes parallel to the substrate plane as
compared to other two samples, shown by peak intensities of fig 3.
Nevertheless, we can analyze the effect of plasma irradiation on grain size
from morphological studies. We have shown such an analysis explicitly in fig
7. Quite good linear behavior shows a definite role of plasma irradiation
on grain size modification. It should be noted that the grain size of as
grown or unirradiated sample was maximum ${\rm (\sim 4.8\mu m)}$ and almost
equal to that of sample irradiated at large distance (12cm). This shows that
plasma ions break the grains into smaller sizes depending on its energy. It
is quite well known that the ion energy decreases linearly with distance
from anode. Therefore, a reasonably good linear behavior in fig 7 may not
be surprising and indicates a good processing method to modify grain sizes.
We further analyze the relation between residual stress and grain size as
shown in figure 8. The positive side of the stress is tensile and those
data are taken from the earlier report \cite{PT4} for the purpose of
comparison. The negative side of the plot indicates the present data of
compressive stress. The main difference between the two sides is the plasma
irradiation. A good linear relation between the stress and grain size in the
present study is the direct consequence of plasma processing. As mentioned
earlier, the bigger grains with considerable stress in as grown film break
into smaller grains with reduced stress and better packing due to plasma
irradiation. Possibly, the modification in grain size and its packing
reduces the residual stress. In contrast, the positive side indicates the
development of residual stress due to the grain size growth and its
distribution. Even the morphology supports this. In the present study the
grain size and its distribution is quite uniform unlike earlier report.
Therefore, the linear relationship in the present study further justifies
and supports the other characterization results.

We have also studied the optical absorption of all the samples. A typical
absorption spectrum of one of the samples is shown in fig 9 for the purpose
of illustration. The absorption coefficient $\alpha$ was calculated as a
function of incident photon energy near the band edge from these absorption
data as explained by Seeger\cite{see}. In general, the relation of the type 
\cite{see, clark} 
\begin{eqnarray}
\alpha h \nu = (h \nu - E_g)^n
\end{eqnarray}
was fitted to the experimental data. The best fit was obtained for n= 1/2
indicating the direct allowed type of transition. The value of E$_g$ is 
determined by the extrapolation of the linear portion in 
the ($\alpha$h$\nu$)$^{2}$ vs h$\nu$ plot as shown in the inset of fig 9. 
Obviously a direct type of transition is evident from the present data.
However, lot of data and discussions are available in the literature on this
issue. Both experiments\cite{3.3, 3.4} and theoretical energy band structure
calculations \cite{band1, band2, band3, band4} reveal the existence of a
direct and indirect band gaps in ${\rm CdI_2}$ differing by only 0.3-0.6eV.
The detailed analysis of this topic can be found in earlier
report\cite{PT4} and the references cited therein. However, we do not
further elaborate on this issue here. We have found the residual stress
dependent ${\rm E_g}$ in this study. There may be several reasons for the
change in ${\rm E_g}$ of the material. The effect of hydrostatic pressure on
${\rm E_g}$ is well known due to the changing atomic distances in the unit
cell. Therefore, similar effect should be expected from residual stress as
well. We have plotted ${\rm E_g}$ as a function of residual stress in fig 10
to examine such a correlation. Again, the negative side of the stress
represents the present data while positive side is taken from the refence
\cite{PT4}. However, it should be noted that we have subtracted a constant
0.5eV from the positive side data for the purpose of matching them at zero
stress. A shift of 0.5eV will not alter the basic trend or the qualitative
behavior of ${\rm E_g}$ with stress.

A reasonably good linear behavior of ${\rm E_g}$ with stress indicates a
definite cause of stress on ${\rm E_g}$. The sharp decrease with tensile
stress has been discussed in the earlier report \cite{PT4}. The overall
linear behavior encompassing both positive and negative sides of stress may
be indicating a common reason for the ${\rm E_g}$ modification. The amount
of change of ${\rm E_g}$ with pressure is given by the slope and is equal to
${\rm 15\mu eV/atm}$. There is only one report\cite{excite} on experimental
study regarding hydrostatic pressure effects on optical excitations in
cadmium halides in the range 1000-3500atm. They have found a linear increase
of ${\rm 5.5\mu eV/atm}$ for an optical edge at 4eV in ${\rm CdI_2}$. The
present data are also in the similar range and show a bit higher rate of
change. The agreement can still be treated as good. However, the little
discrepancy between the two could be due to the anisotropic nature of the
layered structure of ${\rm CdI_2}$. For example, even in the present study
${\rm E_g}$ of (002) and (110) oriented films has different slopes with
pressure. We have shown ${\rm E_g}$ of (110) orientation by triangles in
figure for comparison. The earlier report \cite{excite} was on single
crystal and direction of applied pressure was not mentioned. A consistent
linear behavior with both compressive and tensile stress may well be
indicating a common mechanism of ${\rm E_g}$ change. Referring to electronic
structure calculations of cadmium halides \cite{excite}, the calculated and
experimental band gaps decrease with increasing anion-anion distance in
${\rm CdCl_2}$ to ${\rm CdI_2}$. The exact modification in band structure
due to the fractional change in anion-anion distance is hard to estimate at
present. Therefore, we feel that the changing I-I distance in ${\rm CdI_2}$
due to the residual stress is responsible for the observed changes in ${\rm
E_g}$ with pressure. The amount of change in I-I distance may be different
in differently oriented films and may be different as observed here. The
present data along with earlier similar one \cite{PT4} seem to be very much
consistent and decisive. We hope this will be quite useful in understanding
the interesting layered structured materials like ${\rm CdI_2}$ which has
been quite diverging. Although many experimental results \cite{shin} clearly
show anisotropic optical properties of ${\rm CdI_2}$ the calculated band
structures \cite{band1, band2, band3, band4} show very little or absence of
anisotropic nature.

\section{Conclusions}
The present study highlights the usefulness of plasma processing of ${\rm
CdI_2}$ films in modifying its structural and morpholgical properties. A
good linear relationship between ion energy (or irradiation distance) and
material parameters like cell parameter, grain size, residual stress etc.
indicates the processing as a good technique. The consequent changes ${\rm
E_g}$ correlates well with the residual stress in the film. The combined
data of ${\rm E_g}$ versus compressive and tensile residual stress show a
consistent trend hinting the possible cause as changing iodine-iodine
distance in the unit cell. This could be a useful, decent information and
feedback for revisting band structure calculations of ${\rm CdI_2}$ which is
otherwise diverging.

\begin{acknowledgements}
Authors are thankful to the National Institute of Education, Singapore, for
providing the ARF grant RP 17/00/RSR to fund the research project under
which this investigation has been performed. The discussions with Pankaj
Tyagi is also acknowledged.
\end{acknowledgements}

\pagebreak

\pagebreak

%%%%%%%%%%%%%%%%%%%%%%%%%%%%%%%%%%%%%%%%%%%%%%
%  Figure inclusion (at the end of paper)
%%%%%%%%%%%%%%%%%%%%%%%%%%%%%%%%%%%%%%%%%%%%%%
\begin{figure}
\begin{center}
\epsfig{file=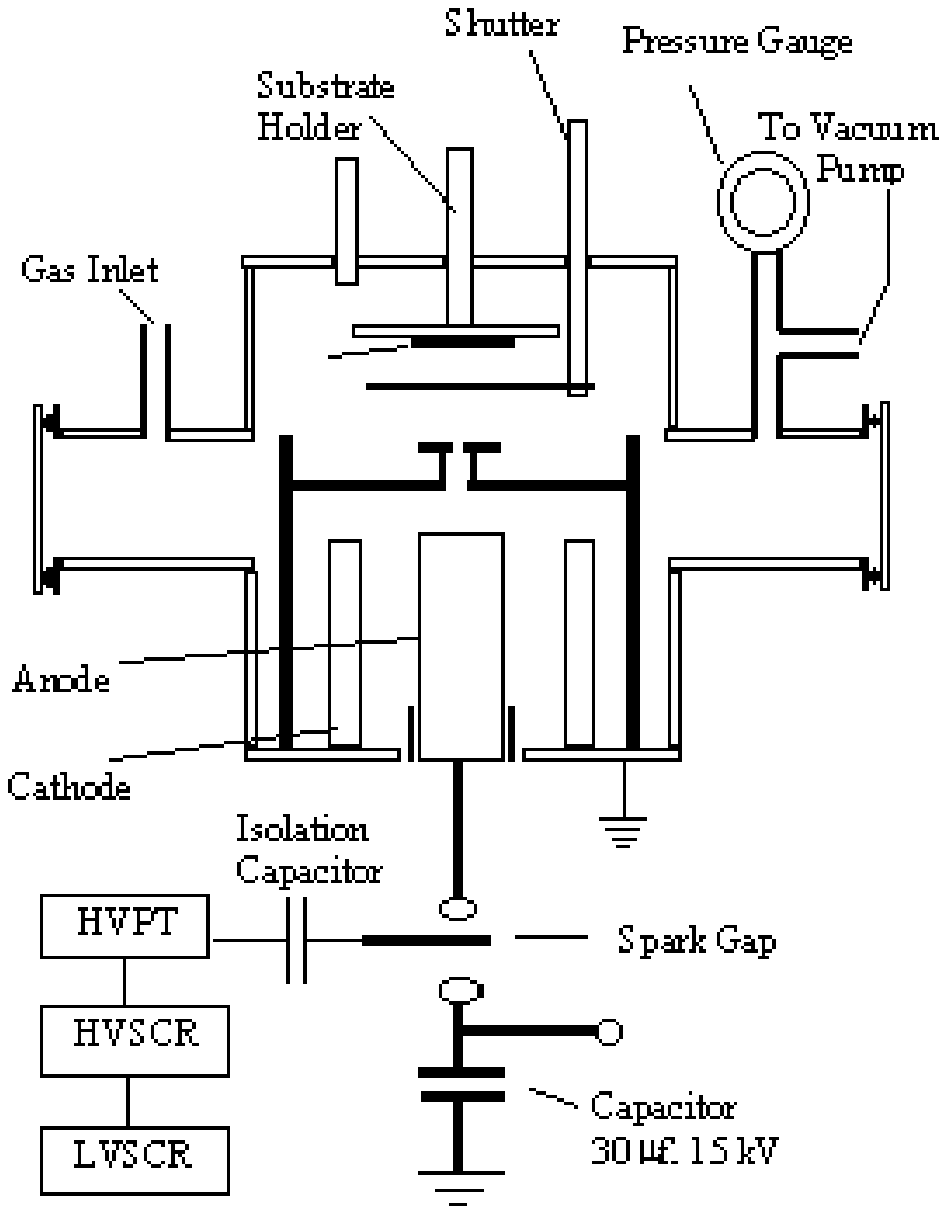,width=5 in}
\caption{ Schematic of experimental set-up for dense plasma focus device.} 
\label{dpf}
\end{center}

\end{figure}
\begin{figure}
\begin{center}
\epsfig{file=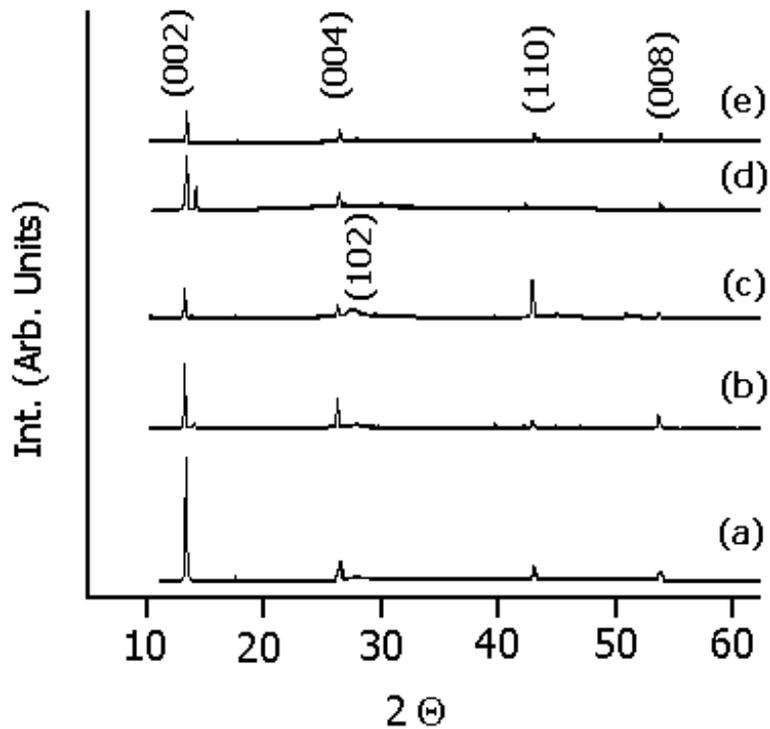,width=4.5 in}
\caption{X-Ray diffractogram of (a) asgrown ${\rm CdI_2}$ film, and films
plasma irradiated at (b) 5cm, (c) 7cm, (d) 10cm, (e) 12cm from the anode.} 
\label{diffracto}
\end{center}
\end{figure}

\begin{figure}
\begin{center}
\epsfig{file=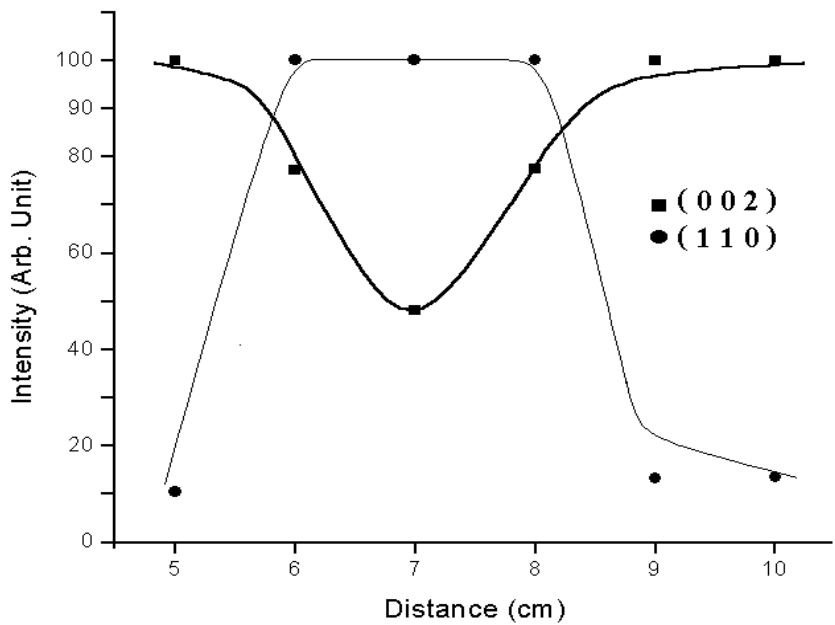,height=3.35in, width=4.5in}
\caption{Variation in intensity of two diffracting planes (002) and (110)
with changing irradiation distance of film from DPF device anode.} 
\label{Intensity}
\end{center}
\end{figure}

\begin{figure}
\begin{center}
\epsfig{file=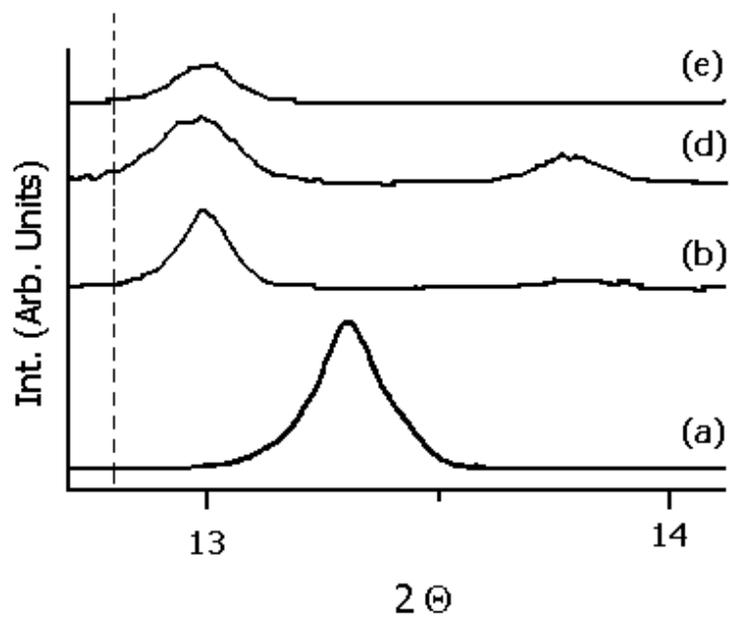, width=4 in}
\caption{Shifts in the (002) peaks of the X-Ray diffractograms shown in
figure(\ref{diffracto}). The labelling are the same as in 
figure(\ref{diffracto}).}
\label{shift}
\end{center}
\end{figure}

\begin{figure}
\epsfig{file=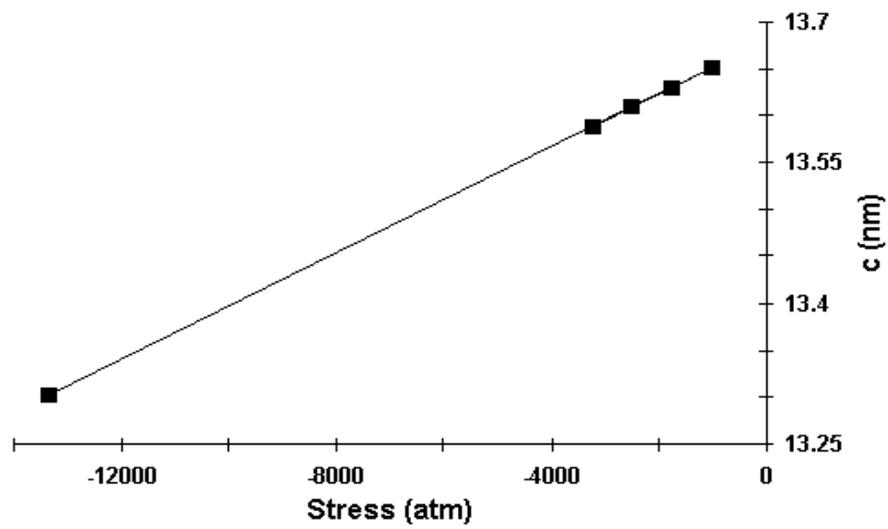,width=5in, height=3in}
\caption{Variation of residual stress with lattice parameter 'c'.}
\label{cstress}
\end{figure} 

\begin{figure}
\begin{center}
\epsfig{file=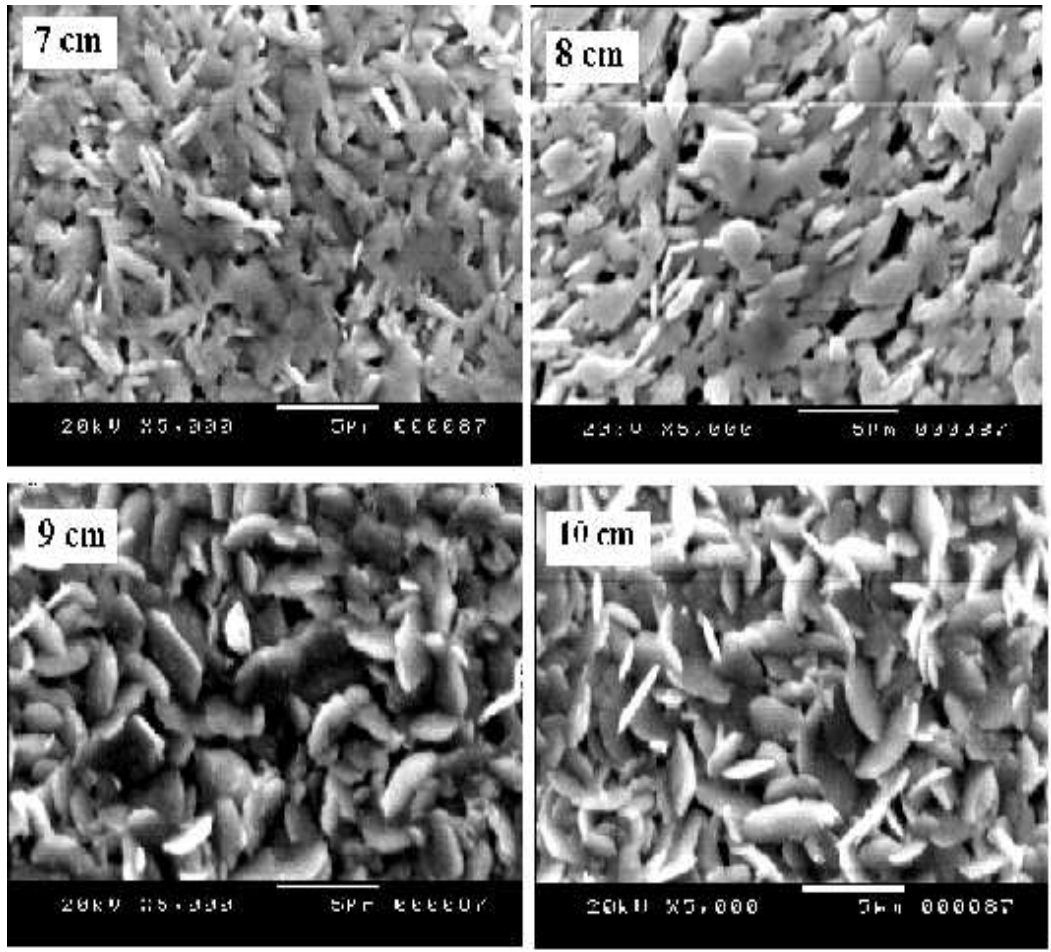,height=5in, width=5.5 in}
\caption{ SEM micrographs of ${\rm CdI_2}$ films irradiated between 7 to
10cm from the anode.}
\label{sem}
\end{center}
\end{figure}

\begin{figure}
\begin{center}
\epsfig{file=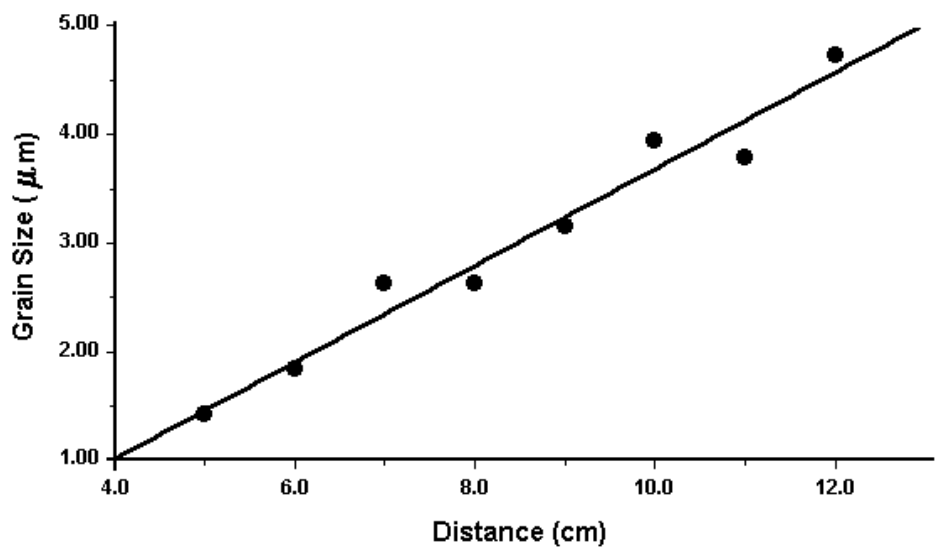, width=5 in}
\caption{Variation of grain size with irradiation distance.}
\label{gsdis}
\end{center}
\end{figure}

\begin{figure}
\begin{center}
\epsfig{file=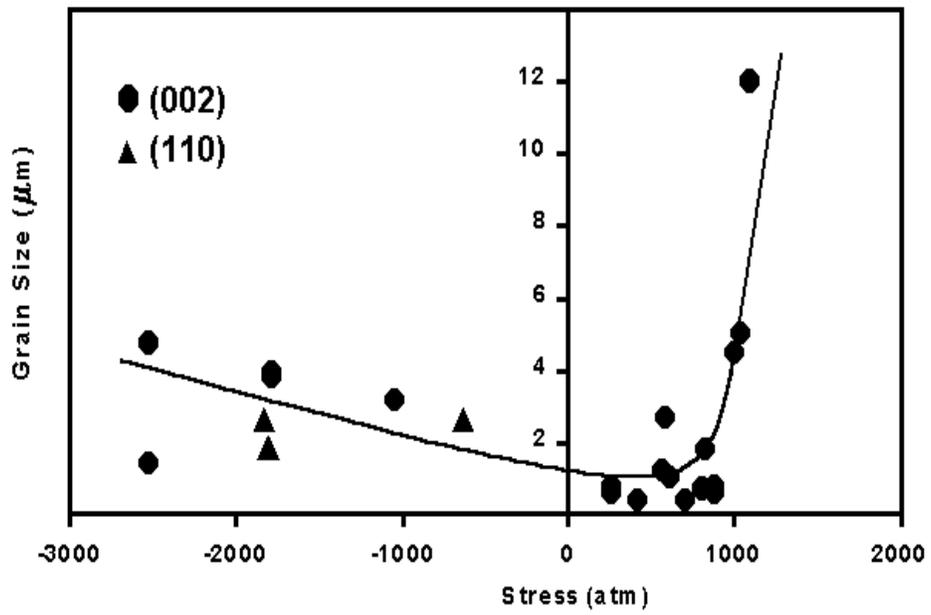, height=3.5in, width=5 in}
\caption{Variation of grain size with residual stress in ${\rm CdI_2}$ films. 
The 'y' axis separates region of compressive and tensile residual stress.}
\label{stressgs}
\end{center}
\end{figure}

\begin{figure}
\begin{center}
\epsfig{file=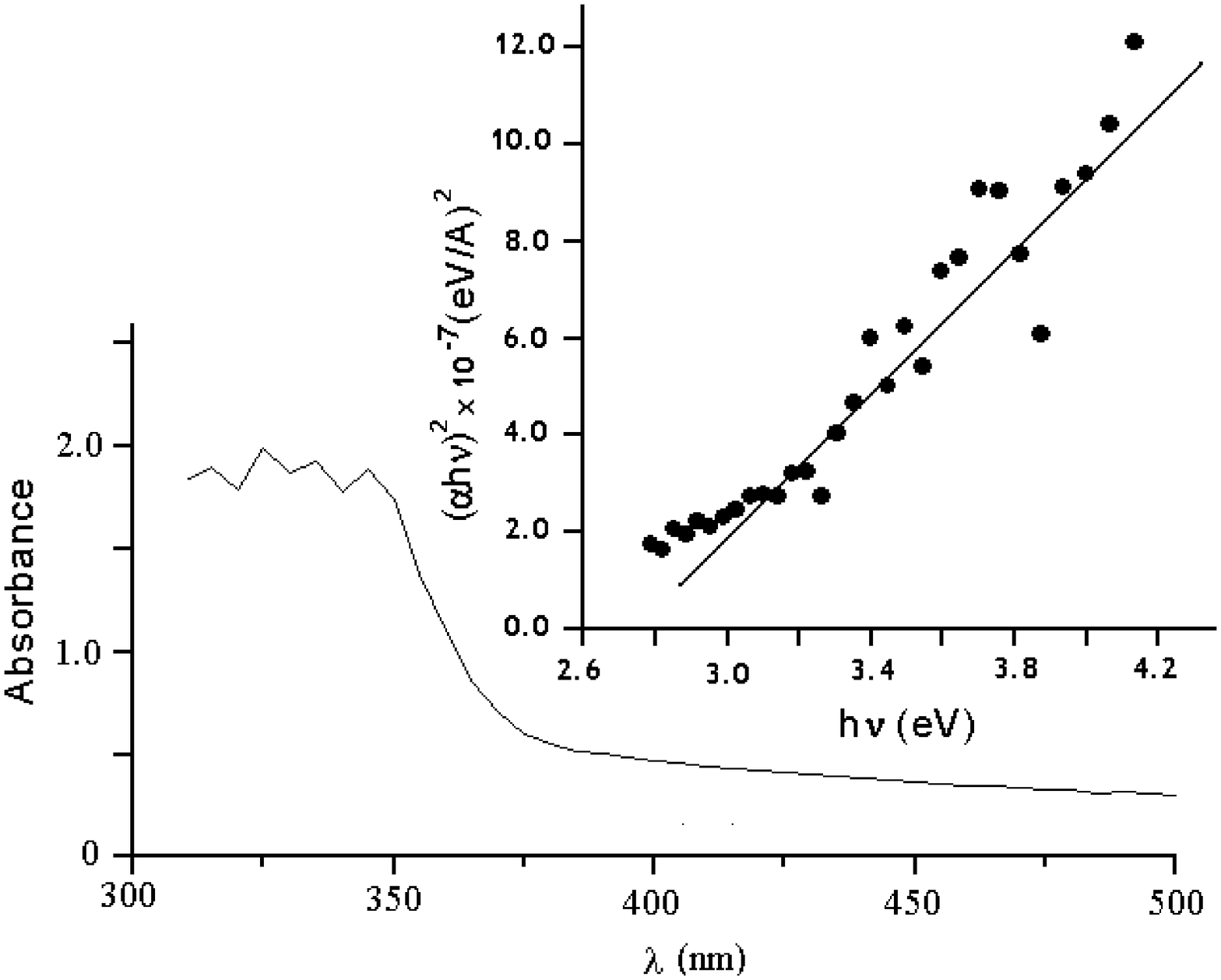, width=5 in}
\caption{ The optical absorption spectra of ${\rm CdI_2}$ film irradiated at
6cm from the anode. The inset shows the fitting of absorption data to eqn 2
with n=1/2 for determining the optical energy gap ${\rm E_g}$.}
\label{uvcdi2}
\end{center}
\end{figure}

\begin{figure}
\begin{center}
\epsfig{file=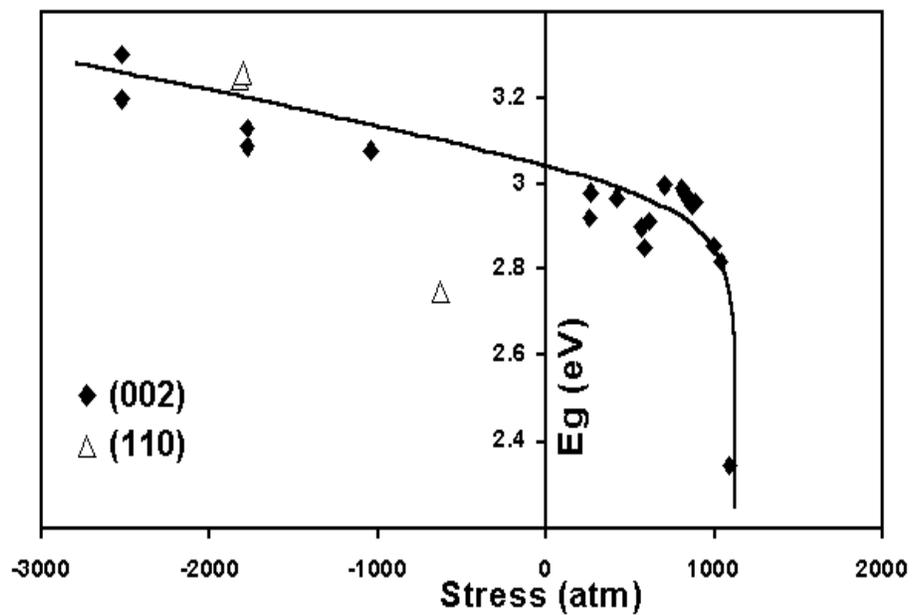,height=3.5in, width=5 in}
\caption{Variation of band gap (${\rm E_g}$), with residual stress for 
${\rm CdI_2}$ films.}
\label{egstress}
\end{center}
\end{figure}

\end{document}